\documentclass[11pt, letter]{article}
\usepackage{jheppub}
\usepackage[english]{babel}
\usepackage{blindtext}\blindmathtrue
\usepackage[dvipsnames]{xcolor}
\usepackage{xparse}
\usepackage{amsmath,epsf,amssymb,mathtools,latexsym,amsthm,setspace,array,pifont,hyperref,amsfonts,dsfont,cancel,braket,parskip}
\usepackage{slashed}
\usepackage{bbm}
\usepackage{graphicx}
\usepackage{tikz}
\usepackage{tikz-feynman}
\tikzfeynmanset{compat=1.0.0}
\usepackage{url}
\usepackage[title]{appendix}

\usepackage[force]{feynmp-auto}
\usepackage{float}
\def\be{\begin{equation}}
	\def\ee{\end{equation}}
\def\beA{\begin{align}}
\def\eeA{\end{align}}

\newcommand{\Lcal}{\mathcal{L}}
\newcommand{\Dcal}{\mathcal{D}}
\newcommand{\Rcal}{\mathcal{R}}
\newcommand{\Ocal}{\mathcal{O}}
\newcommand{\Ucal}{\mathcal{U}}

\newcommand{\Tr}{\text{Tr}}
\newcommand{\tr}{\text{tr}}

\def\prd{\ref@{Phys.~Rev.~D}}        

\title{$\eta$ regularisation and the functional measure}

\author{Robert G. C. Smith [1]}
\author{and Murdock Grewar [2]}

\affiliation{[1]School of Physics and Astronomy, University of Nottingham, University Park, Nottingham NG7 2RD, United Kingdom}
\affiliation{[1]The Nottingham Centre of Gravity, University of Nottingham, Nottingham NG7 2RD, UK}
\affiliation{[2]Department of Materials Physics, Research School of Physics, Australian National University, Canberra ACT 2600, Australia}

\emailAdd{robert.smith@nottingham.ac.uk}
\emailAdd{murdock.grewar@anu.edu.au}
\abstract{In this paper, we revisit Fujikawa’s path integral formulation of the chiral anomaly and develop a generalised framework for systematically defining a regularised functional measure. This construction extends the $\eta$ regularisation scheme \citep{PadillaSmith1,PadillaSmith2} to operator language, making the connection between spectral asymmetry and measure transformation fully explicit. Before recovering Fujikawa’s expression for the chiral anomaly from the regularised measure, we explore the deeper number-theoretic structure underlying the ill-defined spectral sum associated with the anomaly, interpreting it through the lens of smoothed asymptotics. Our approach unifies two complementary perspectives: the analytic regularisation of Fujikawa and the topological characterisation given by the Atiyah–Singer index theorem. We further investigate how the measure transforms under changes to the regularisation scale and derive a function $\iota_E(\Lambda)$ that encodes this dependence, showing how its Mellin moments govern the appearance of divergences. Finally, we comment on the conceptual relationship between the regularised measure, $\eta$ regularisation, and the generalised Schwinger proper-time formalism, with a particular focus on the two-dimensional Schwinger model.}

\begin{document}
\maketitle

\section{Introduction}
The path integral formulation of quantum field theory (QFT) provides a powerful framework for analysing the dynamics, symmetries, and topological properties of quantum fields \cite{Das:1993gd,Rivers1987PathIM, Bastianelli:2006rx}. By defining the path integral as a summation of all possible field configurations, weighted by the exponential of the classical action, it elegantly unifies quantum dynamics with classical principles, realising Feynman’s original vision \cite{Feynman1965QuantumMA}. One notable advantage of this approach over operator-based methods \cite{Bastianelli:2006rx} is the ability to make the connection between symmetries, conservation laws, and quantum anomalies transparent \cite{Fujikawa:2004cx}. Its ability to seamlessly incorporate symmetries and constraints through functional determinants underscores its indispensability in the study of gauge theories \cite{Fujikawa:2004cx}, a topic central to modern physics. The path integral formalism is also particularly effective for exploring non-perturbative phenomena \cite{Das:1993gd}, where canonical methods can prove cumbersome or impractical. 

Rigorously defining path integrals in physically relevant QFTs poses formidable technical difficulties \cite{Glimm:1987ylb}. Many of these difficulties arise from issues related to the definition of the integration measure \cite{Glimm:1987ylb, Bastianelli:2006rx}. In standard quantum mechanics, this is less problematic, as the path integral can be rigorously formulated using measures defined on abstract Wiener spaces \cite{Hall2013}. In QFT, however, similar attempts to rigorously define the measure have led to approaches such as the Osterwalder-Schrader axioms \cite{Glimm:1987ylb, Osterwalder:1973dx, Osterwalder:1974tc}. Although progress has been made, issues such as the convergence properties of integrals remain unresolved, and a fully rigorous definition of the measure in the general case remains elusive.
 
Although these are primarily mathematical issues, they are far from irrelevant in physics. One area where considerations about the path integral measure play an important role is in the study of anomalies. In QFT, classical symmetries are often broken by quantum effects. A quantum anomaly is defined as the failure of a classical symmetry to be preserved in the process of quantisation and regularisation. In the case where the classical symmetry is a global one, the breaking of this symmetry is not necessarily problematic \cite{Kallosh:1995hi, Bastianelli:2006rx, Branchina:2024xzh, Branchina:2024lai}. Gauge anomalies, on the other hand, can be fatal, as they signal an inconsistency in the quantum theory. They lead to the breakdown of gauge invariance, which can cause violations of unitarity and introduce unphysical degrees of freedom in the effective field theory. In the path integral formulation, these gauge anomalies reveal subtle inconsistencies with classical symmetries in the functional measure when quantised \cite{Fujikawa:2004cx}.

The prototypical playground for studying anomalies is chiral gauge theory in an even number of spacetime dimensions, in which a gauge field is coupled to massless Dirac fermions \cite{Bertlmann:1996xk}. Classically, there are two conserved currents: a vector current associated with the gauge symmetry and an axial current associated with the chiral symmetry. In two dimensions, perturbative calculations show there are cases for certain quantum field theories where two point functions reveal both symmetries can be anomalous, although somewhat unusually, this does not necessarily lead to any inconsistencies \citep{Jackiw:1984zi}. In other cases, depending on the structure of the theory, the chiral anomaly is a genuine gauge anomaly, meaning that it obstructs gauge invariance. 

Furthermore, in four dimensions, if the chiral anomaly affects a global axial symmetry (like the baryon number in QCD or the axial symmetry in the Standard Model), it does not necessarily signal an inconsistency. If, on the other hand, the chiral anomaly were to affect a gauge symmetry (e.g., if a gauge theory has chiral fermions that lead to a gauge current anomaly), then it would be a gauge anomaly, and the theory would suffer from inconsistencies. In the simplest setting \citep{Adler:1969gk, Bell:1969ts} for the perturbative computation of the decay processes of a neutral pion to two photons, the three point functions in four dimensions revealed such an anomalous chiral current. However, the underlying gauge symmetry remained intact.

In all these cases of taking a perturbative approach to anomaly calculations, a key feature is the need for regularisation. The definition of chiral models makes use of the $\gamma_5$ matrix, which, for certain regularisation schemes like dimensional regularisation, is the source of significant ambiguities \cite{Breitenlohner:1977hr,Ferrari:2014jqa,Ferrari:2015mha,Tsai:2010aq,Tsai:2009it}. Specifically, in $d \neq 4$ dimensions, $\gamma_5$ is not uniquely defined, leading to multiple prescriptions for handling its anticommutation relations with other Dirac matrices. In general non-perturbative approaches, $\gamma_5$ can also cause ambiguities in operator ordering, while, in Fujikawa's path integral approach \citep{Fujikawa:2004cx}, a major subtlety in the spectral analysis has to do with $\gamma_5$ not being a trace-class operator.

The path integral approach developed systematically by Fujikawa \cite{Fujikawa:1979ay, Fujikawa:1980eg} and later by Fujikawa and Suzuki \cite{Fujikawa:2004cx} to studying chiral anomalies, reveals classical symmetry breaking during quantisation much more explicitly than in perturbative approaches. In fact, Fujikawa’s analysis has become fundamental in modern field theory, as it demonstrates the way in which anomalous effects of symmetry breaking arise directly from a non-trivial Jacobian associated with a change of variables in the path integral. A key observation made by Fujikawa is that the path integral measure itself breaks certain classical symmetries. Due to its explicit nature, the Fujikawa path integral method provides a conceptually clear framework for deriving anomalous identities and examining the effects of regularisation. This approach allows anomaly terms and associated ultraviolet (UV) divergences to appear early in the calculation, offering a transparent view of their origins.

Fujikawa's approach identifies the quantum anomaly as the determinant of the transformation in the space of paths, which for anomalous symmetries is nonzero. However, to compute these determinants, a regulator is required to ensure the calculation converges. This regulator is not a part of the original formulation of the path integral formalism, but something which is added on an ad-hoc basis at the time the anomaly is computed.\footnote{In addition to Fujikawa's regularisation, time-slicing and zeta function regularisation are also commonly used \citep{Bastianelli:2006rx}. Alternatively, proper-time regularisation is also used, which is closely related to the heat kernel regularisation and Fujikawa's smooth cutoff regularisation \citep{Ball:1988xg, Leutwyler:1985ar, Umezawa:1989ns}.} It is not clear, \emph{a priori}, that anomalies computed in this way would be consistent with one another, as they are not derived from an initial regularised formulation of the path integral.

If one starts from the view that, roughly speaking, integration in the Feynman path integral is integration with respect to a pseudomeasure \cite{johnson2000feynman, cartier2006functional}, the primary issue is that one cannot construct the Feynman path integral as used in physics as a mathematical representation of a standard Lebesgue integral \citep{stroock2000introduction,janke2008path}. There are multiple ways to circumvent this issue, with each having its own advantages and suffering its own drawbacks. But any mathematical attempts at formalising a generalised measure must also address the need for regularisation in practical calculations. This raises an interesting question: how might one more rigorously construct a regularised path integral, as understood in its physics application?  

For example, in \cite{Montaldi2016FeynmanPI} a formalisation of Feynman's original definition is considered, with the formal generalised Lebesgue-Feynman measure confirming the origin of quantum anomalies as given by Fujikawa and Suzuki \cite{Fujikawa:2004cx}. However, the general role of renormalisation in the calculation of quantum anomalies was left open. Most recently \cite{costello2022}, and perhaps most notably, work has been done at the foundations of perturbative QFT following the Wilsonian philosophy of renormalisation. Here, it is proposed that instead of attempting to define integrals over infinite-dimensional space such as $C^{\infty}(M)$, one should instead calculate path integrals over a finite-dimensional approximation $C^{\infty}(M)_{\leq \Lambda}$ of the infinite-dimensional space, using an energy cutoff scale $\Lambda$. The motivation, which we note is somewhat similar to the heat kernel approach in \citep{Fujikawa:2004cx}, is closer to the intuition utilised in practical models in physics, where one employs a regularisation scheme typically in the form of a high energy cutoff.

In this work we seek to answer the previously stated question, reproducing the chiral anomaly calculation in Quantum Electrodynamics (QED) by formulating the theory with a functional measure that is regularised from the outset. Our approach is inspired by the generalised $\eta$ regularisation developed on the level of perturbative loop integrals in \cite{PadillaSmith1, PadillaSmith2}, where smooth regulator functions enter as a modification of the integration measure. We show that in our approach to defining a regularised path integral measure, we effectively extend the notion of $\eta$ regularisation for loop integrals to the functional integral. Our framework is shown to reproduce Fujikawa's results for the chiral anomaly in a general sense, capturing his original regularisation scheme. We also show how our generalised framework relates to recent efforts that calculate the chiral anomaly in two dimensions using $\eta$ regularisation for perturbative loop integrals, while also commenting on the relation to the Schwinger proper time formalism.

The rest of this paper is organised as follows: in the next section we review Fujikawa's general formalism and calculation of the chiral anomaly. We emphasise the way in which the chiral anomaly may be formally related to an infinite (divergent) series of the form $1+1-1-1+\dots$, and show how for Fujikawa's choice of regulator the smoothed asymptotic method of $\eta$ regularisation reveals this series to converge. We then discuss the physical meaning of this as it relates to the arrangement of eigenvalues in Fujikawa's path integral calculation. In Section 3, we introduce our generalised framework and define the regularised measure. We show that our formalism reproduces the regularised sum in the same way as Fujikawa's method, except that this regularised sum follows as an automatic consequence of regularising the functional measure from the outset. We then go on to show how the regularised measure may be related to the ``enhanced" regulators and the condition of vanishing Mellin transforms of the $\eta$ regularisation scheme \cite{PadillaSmith1}. Finally, in Section 4, we study the chiral anomaly for the two dimensional Schwinger model, and show that for the appropriate choice of regulator in the regularised measure we recover the same expression for the trace over Dirac indices as recently computed for $\eta$ regularisation in \citep{PadillaSmith2}, offering further connection between the different frameworks.

\section{A review of Fujikawa's path integral method and the chiral anomaly}\label{Fujikawa}
To review Fujikawa's path integral method in the study of the chiral anomaly, we follow \cite{Fujikawa:1979ay,Fujikawa:1980eg,Fujikawa:2004cx} and the conventions defined therein. We begin with the full Euclidean path integral for QED, incorporating both the Maxwell and Dirac actions as well as gauge-fixing through the Faddeev-Popov procedure. The generating functional in Euclidean space is given by

\be\label{FullQEDPartitionFunct}
Z[J_\mu, \bar \xi, \xi] = \int \mathcal{D}\Phi \, \exp \left\{ -\frac{1}{\hbar} \int d^4x \left( \mathcal{L}_\text{QED} + J_\mu A^\mu + \bar{\xi}\psi + \xi \bar{\psi} \right) \right\},
\ee
where 
\be
\Lcal_{\text{QED}} = \left( \psi \gamma^\mu (\partial_\mu - e A_\mu) \psi - m \bar\psi \psi - \frac{1}{4} F_{\mu \nu} F^{\mu \nu} \right)
\ee
with $F_{\mu \nu} = (\partial_\mu A_\nu - \partial_\nu A_\mu)$ and source terms denoted as $J_\mu$ and $\xi$. The extended field measure $\Dcal \Phi$ is defined to include the terms as a result of the Faddeev-Popov gauge-fixing procedure to handle gauge redundancy. It is given by
\be\label{FullQEDLagrangianFPterms}
\Dcal \Phi = \Dcal\bar{\psi} \Dcal \psi \left( \prod_\mu \Dcal A_\mu \right) \Dcal B \Dcal \bar{c} \Dcal c \exp \left\{ \frac{1}{\hbar} \int d^4x \left( (\partial_\mu \bar{c}) (\partial^\mu c) - (\partial_\mu B) A^\mu \right) \right\}.
\ee

The `auxiliary field' $B$ enforces the Landau gauge $(\partial_\mu A^\mu = 0)$, which is an incomplete gauge-fixing, and the Faddeev-Popov ghosts $\bar c, c$ remove the remaining gauge redundancy \cite{Fujikawa:2004cx}. The form of \eqref{FullQEDPartitionFunct} already implies a rescaling of $A_{\mu}^{\alpha}$ and $F^{\alpha}_{\mu \nu}$ to absorb factors of the gauge coupling constant \cite{Das:1993gd}, resulting in the matter part of the Lagrangian being quadratic in the field strength.

The Euclidean path integral \eqref{FullQEDPartitionFunct} that Fujikawa considers is obtained by a Wick rotation $x^4 = -i x^0$ \cite{Fujikawa:2004cx}. The spacetime metric signature is $g_{\mu \nu} = \left(-1,-1,-1,-1\right)$ and $A_4 = -iA_0$. The operator $\slashed{D} = \gamma^{\mu}(\partial_{\mu} + iA_{\mu})$ becomes a Hermitian operator $\slashed{D} = \gamma^{\mu} D_{\mu}$ in Euclidean space. We also note that for the Clifford algebra, $\gamma^0 \to -i\gamma^4$ and all $\gamma^\mu$ matrices become anti-Hermitian $(\gamma^{\mu})^{\dagger} = -\gamma^{\mu}$, with the anticommutation relation defined as $\{\gamma^{\mu}, \gamma^{\nu} \} = 2 g^{\mu \nu}$. We also note the following conventions:

\be
\gamma_5 = \gamma^4 \gamma^1 \gamma^2 \gamma^3, \quad \gamma_5^2 = 1, \quad \gamma_5^{\dagger} = \gamma_5, \quad \{\gamma_5, \gamma^{\mu} \} = 0.
\ee

Following Fujikawa \citep{Fujikawa:1979ay,Fujikawa:2004cx}, in the study of the chiral anomaly one is not required to give the precise form of the gauge field Lagrangian, and so we set aside discussion on the subtleties of gauge fixing and the precise form of the gauge field propagator. In fact, as Fujikawa showed, most important in the present context is the matter Lagrangian and the matter measure. It is thus sufficient for the purposes of the present discussion to consider the massless theory and suppress the source terms, and we include the Fadeev-Popov factor into $[\Dcal A_\mu]$ whenever necessary. To simplify notation we use natural units $\hbar = 1$. Taking these steps, we are left with the simplified partition function for massive spinor QED as originally considered by Fujikawa \cite{Fujikawa:1979ay}

\be
Z = \int \Dcal A_\mu \Dcal \bar{\psi} \Dcal \psi \, e^{S[A, \bar{\psi}, \psi]},
\ee
with the Lagrangian 
\be\label{FujikawaLagrangian}
\Lcal = \bar{\psi}\left(i\slashed{D} - m \right)\psi -\frac{1}{4} \Tr F_{\mu\nu}F^{\mu \nu}.
\ee

Fujikawa's formalism computes anomalies by examining the Jacobian determinant of the transformation of the field measure; here we follow \citep{Fujikawa:2004cx} in particular. To facilitate the calculation of the determinant, the fermionic fields $\bar \psi, \psi$ are explicitly decomposed as summations over countable bases according to the free modes of propagation, i.e. according to the eigenfunctions of the Dirac operator $\slashed{D}$, viz.
\be
\psi = \sum_n a_n \phi_n(x) , \qquad \bar \psi = \sum_n \bar{b}_n \phi^\dag_n(x) \, .
\ee
The fields $\bar \psi, \psi$ are Grassmann-number-valued, as the $a_n$ and $\bar{b}_n$ are generators of a Grassmann algebra. The $\phi_n(x)$ are the spinor eigenfunctions of $\slashed D$:
\be
\slashed D \phi_n = \lambda_n \phi_n \, .
\ee
The total path integral measure, denoted $\mathrm d \mu$, has its fermionic part decomposed viz.
\be\label{Pathintegralmeasure}
d \mu \equiv \left[\Dcal A_{\mu}(x)\right] \Dcal \bar{\psi} \Dcal \psi = \left[\Dcal A_{\mu}(x)\right] \prod_{n} \mathrm d \bar{b}_n \mathrm d a_n \, .
\ee

Quantum anomalies are symmetry-breaking which is not expected in the classical theory, and Fujikawa's formalism attributes such anomalies to a change in the functional measure. 
A \emph{symmetry} may be defined as a transformation for which the correlation functions involving the transformed variables are unaltered by the transformation, e.g. 
\be
\langle \dots \psi(x) \dots \rangle = \langle \dots \psi'(x) \dots \rangle \, .
\ee

In the path integral formalism, this definition turns out to be equivalent to the requirement that the product of the measure and the integrand of the generating functional be invariant when the source terms $J, \bar{\xi}, \xi = 0$. In the case of the QED generating functional in \eqref{FullQEDPartitionFunct}, this amounts to the requirement that:
\be\label{eq:cadf_equality}
\Dcal \bar{\psi} \Dcal \psi \, \exp \left\{ \int d^4x \mathcal{L}_\text{QED} \right\}
= \Dcal \bar{\psi}' \Dcal \psi' \, \exp \left\{ \int d^4x \mathcal{L}_\text{QED}' \right\} \, .
\ee

To study the symmetry properties of a local transformation $U(x)$ of the spinors $\bar{\psi}, \psi$, the transformed variables are introduced:
\be\label{LocalTransformations}
    \psi'(x) = U(x) \psi(x) \, , \quad \bar{\psi}'(x) =\bar{\psi}(x) \bar{U}(x) \quad \text{where} \quad \bar{U}(x) = -\gamma^4 U^\dag(x) \gamma^4 \, ,
\ee
where the transformation for the adjoint spinor $\bar{\psi}(x) = \psi^{\prime}(x)^{\dagger}(-i\gamma^4)$ and the standard identity $\gamma^4 \gamma^4 = -1$ is used. As can be seen, this implies the relation $\bar{\psi}(x) \rightarrow \bar{\psi}^{\prime}(x) = \psi^{\dagger}(x)(-i\gamma^4)(-\gamma^4 U^{\dagger}(x)\gamma^4) = -i \psi^{\prime}(x)^{\dagger}\gamma^4 U^{\dagger}(x)$.\footnote{If we had considered the full partition function \eqref{FullQEDPartitionFunct} with the full functional measure derived using the Fadeev-Popov procedure, general local transformations would require further information.}
In the case of the chiral anomaly, the transformation is defined by $U(x) = e^{i \gamma^5 \alpha(x)}$, which gives $\bar{U}(x) = U(x) = e^{i \gamma^5 \alpha(x)}$, and the transformed variables:
\be
    \psi'(x) = e^{i \gamma^5 \alpha(x)} \psi(x) \, , \quad \bar{\psi}'(x) = \bar{\psi}(x) e^{i \gamma^5 \alpha(x)} \, .
\ee

In operator form, the same equation reads
\be
    \langle x | \Ucal |\psi\rangle = \langle x | e^{i \gamma^5 \alpha(x)} | \psi\rangle \, , \quad \langle\bar{\psi}| \bar{\Ucal} | x \rangle = \langle\bar{\psi}| e^{i \gamma^5 \alpha(x)} | x \rangle \, .
\ee

The transformed Lagrangian $\mathcal L_{\text{QED}}'$ (which is obtained from $\mathcal L_{\text{QED}}$ by performing the substitution $\psi \mapsto \psi', \bar{\psi} \mapsto \bar{\psi}'$) is unchanged---this means that the chiral transformation is classically a symmetry. The quantum anomaly, which is the breaking of the classical symmetry when the theory is quantised, arises because the equality \eqref{eq:cadf_equality} does not hold, despite the Lagrangian on either side being equal. The inequality is down to the change in measure, i.e. the following measures are not the same:
\be
    \Dcal \bar{\psi} \Dcal \psi \neq \Dcal \bar{\psi}^{\prime} \Dcal \psi^{\prime} \, .
\ee

As a matter of notation, the measure on the left side of the equation is understood in a consistent manner with \eqref{eq:cadf_equality}. Therefore, the transformed fields are expanded in the basis of eigenfunctions $\phi_n(x)$ of $\slashed D$ and expressed as a linear combination with coefficients given by Grassmann generators,
\be
\psi' = \sum_n a_n' \phi_n , \qquad \bar \psi' = \sum_n \bar{b}_n' \phi^\dag_n \, ,
\ee
and the measure is given by
\be\label{eq:primed_measure}
    \Dcal \bar{\psi}' \Dcal \psi' 
    = \prod_{n} \mathrm d \bar{b}_n' \mathrm d a_n' \, .
\ee

This may be related to the original measure $\Dcal \bar{\psi} \Dcal \psi$ by expanding both sides of the equations $\psi' = \Ucal \psi$ and $\bar{\psi}' = \bar{\psi} \bar{\Ucal}$ viz.
\begin{align}
    \sum_n a_n' |\phi_n \rangle &= \sum_n a_n \Ucal |\phi_n \rangle \, , &
    \sum_n \bar{b}_n' \langle\phi_n| &= \sum_n \bar{b}_n \langle\phi_n| \bar{\Ucal} \, .
\end{align}

For the first equation, matching the coefficients of $|\phi_n \rangle$ on either side yields $a_n'$ as a combination of the $a_n$. A similar exercise is performed for $\bar{b}_n'$ in terms of $\bar{b}_n$. One finds that
\be
    \bar{b}_n' = \sum_m \langle \phi_m | \bar{\Ucal} | \phi_n \rangle \bar{b}_m = \sum_m \bar{\Ucal}^{\mathrm T}_{nm} \bar{b}_m
    \, , \qquad
    a_n' = \sum_m \langle \phi_n | \Ucal | \phi_m \rangle a_m = \sum_m \Ucal_{nm} a_m \, .
\ee

Since integration over Grassmann generators is defined by differentiation, the resulting Jacobian of the transformation is the inverse of that for conventional numbers. Therefore, using also $\det \bar{\Ucal}^{\mathrm T} \det \Ucal = \det \bar{\Ucal} \det \Ucal = \det \left(\bar{\Ucal} \Ucal \right)$, we find the relation between the measures
\be
    \Dcal \bar{\psi}' \Dcal \psi' 
    = \det \left(\bar{\Ucal} \Ucal \right)^{-1} \Dcal \bar{\psi} \Dcal \psi \, .
\ee

It is now clear to see why the chiral transformation given by $\Ucal = \bar{\Ucal} = e^{i \gamma^5 \alpha(x)}$ results in a change to the measure. Compare with a non-anomalous symmetry such as the gauge symmetry $\Ucal = e^{i \alpha(x)}$, for which $\bar{\Ucal} = e^{-i \alpha(x)}$ and $\bar{\Ucal} \Ucal = 1$; in that case the measures are equal, and so the symmetry is anomaly-free.

The factor $\det(\bar{\Ucal} \Ucal)^{-1}$ is the Jacobian of the transformation. To compute it, we use the identity $\det(\exp(A)) = \exp(\Tr(A))$. The trace is a functional trace which is computed over the Dirac indices and the spatial indices, so it consists of a combined summation and integration. The Jacobian is found to be
\begin{eqnarray}\label{Uchiral2}
(\det e^{2i\gamma^5 \alpha(x)})^{-1} 
&=& \exp\left(-2i \sum \limits_{n=1}^{\infty} \int d^4x \phi^{\dagger}_n(x) \gamma_5 \alpha(x) \phi_n(x) \right).
\end{eqnarray}

The right-hand side is not well-defined because the operator $\bar{\Ucal} \Ucal = e^{2i \gamma^5 \alpha(x)}$ is not \emph{trace-class} \cite{conway2000course,reed1980methods}, which is to say it is not a compact operator on the Hilbert space whose singular values are summable (ensuring the trace is finite and well-defined). The anomaly is not well-defined until/unless the summation is regulated. In the next section, we will focus on the explicit structure of this sum in more detail. Important at this juncture is how Fujikawa resolves the ill-definedness of the summation by applying a regularisation scheme. The regulated form of the functional trace in \eqref{Uchiral2} is defined in terms of a smooth regulating function $f : \mathbb R \rightarrow \mathbb R$ as follows\footnote{There are several reasons for why this choice of regularisation is natural in the context of the functional trace \eqref{Uchiral2}, with the method further generalised by Umezawa \citep{Umezawa:1989ns}. Most importantly, $f(\slashed{D}^2/\Lambda^2)$ is clearly gauge invariant.As explained in \citep{Weinberg95a,Bilal08}, although $f\left(D_{\mu} D^{\mu}/\Lambda^2\right)$ is also gauge invariant, and would result in no $\gamma$ matrices being contained in the regulator and thus a vanishing quantity in the sum, the requirement that one must not only regularise the determinant but also the fermion propagator means the regulator function cannot be taken to be a function of $D_{\mu}D^{\mu}$. Moreover, since as we are technically dealing with fermions in an external gauge field $A$, the appropriate space on which we should compute the sum (or functional trace) is spanned by the eigenfunctions of $\slashed{D}$ and not $D_{\mu}D^{\mu}$. On the other hand, the choice $f\left(\slashed{\partial}^2/\Lambda^2 \right)$ would break gauge invariance.}

\begin{equation}\label{Uchiral2FujiReg1}
\lim_{\Lambda \rightarrow \infty} \! \exp\left( \Tr(f(\slashed D^2 / \Lambda^2) 2i \gamma^5 \alpha(x)) \right) 
=
\lim_{\Lambda \rightarrow \infty} \! \exp\left(-2i \sum \limits_{n=1}^{\infty} \int d^4x \phi^{\dagger}_n(x) f(\slashed D^2 / \Lambda^2) \gamma_5 \alpha(x) \phi_n(x) \right) \, .
\end{equation}

The smooth function $f$ applied to an operator, i.e. $f(\slashed D^2 / \Lambda^2)$, is understood in the usual sense of functional calculus. Furthermore, $f$ is required to satisfy $f(0) = 1$ and $\lim_{x\rightarrow \infty} f(x) = 0$. Fujikawa computes the regularised trace for an arbitrary choice of $\alpha(x)$, which involves lengthy and subtle computation \cite{Fujikawa:2004cx}. The result is independent of $f$, provided $f(x)$ decays sufficiently quickly with $x$. The result is 

\begin{equation}\label{Uchiral2FujiReg2}
\lim_{\Lambda \rightarrow \infty} \exp\left( \Tr(f(\slashed D^2 / \Lambda^2) 2i \gamma^5 \alpha(x)) \right) 
= \exp\left(-2i \int d^4x \, \alpha(x) \frac{e^2}{32\pi^2}\epsilon^{\mu \nu \alpha \beta}F_{\mu \nu}F_{\alpha \beta}  \right) \, .
\end{equation}

This agrees with the result for the chiral anomaly originally found by Adler, Bell, and Jackiw \cite{Adler:1969gk, Bell:1969ts}. It is straightforward to see \cite{Fujikawa:2004cx} that when deriving the Ward identities, the chiral (axial) current $j_A^\mu = i\bar{\psi}\gamma^\mu \gamma^5 \psi$ associated with axial transformations is no longer conserved $\partial_\mu j^\mu_A = e^2/(16\pi^2) \epsilon^{\mu \nu \alpha \beta} F_{\mu \nu}F_{\alpha \beta}$. Similar is true in the case of massive fermions with the exception that, classically, the axial current is not conserved $\partial_\mu J_5^\mu = 2i m \bar{\psi} \gamma^5 \psi$. The axial anomaly still contributes at the quantum level, with the non-conservation of the axial current taking the form $\partial_\mu J_5^\mu = 2i m \bar{\psi} \gamma^5 \psi + e^2/(16\pi^2) \epsilon^{\mu\nu\alpha \beta} F_{\mu\nu} F_{\alpha \beta}$. In both cases, we see that the total charge is $e$ times a topological invariant.

\subsection{Spectral summation and smoothed asymptotics}\label{EigensumSeries}
In this section, we illustrate how the ill-definedness of the sum in \eqref{Uchiral2} manifests when one tries to compute the chiral anomaly. In particular, we examine how the spectral summation \citep{Chamseddine:1998wq,Hall2013QuantumTF,Moretti:2013cma,Fujikawa:2022cee} connects to smoothed asymptotics  \citep{PadillaSmith1} from a number theory perspective.

We begin by noting that the unregularised sum effectively looks like a trace over the chiral operator $\Tr(\gamma_5)=0$, which gives zero. But as it is also weighted by the eigenfunctions $\phi_n(x)$, the total sum is a trace over an infinite number of states reflecting the infinite degrees of freedom in the definition of the measure \eqref{Pathintegralmeasure}. For the case that there is an eigenfunction $\phi_n(x)$ with eigenvalue $\lambda_n \neq 0$, then $\gamma^5 \phi_n(x)$ is also an eigenfunction \citep{Tonggaugetheory}. The reason for this follows from how $\gamma^\mu \gamma^5 = -\gamma^5 \gamma^\mu \) such that \citep{Tonggaugetheory}

\be
\slashed{D}(\gamma^5 \phi_n) = -\gamma^5 \slashed{D} \phi_n = -\lambda_n \gamma^5 \phi_n,
\ee

as clearly demonstrated by Fujikawa \citep{Fujikawa:2004cx} by taking the basis set $\phi_n(x)$---which is required to diagonalise $\slashed D$---to simultaneously diagonalise $\gamma_5$. The key observation is that all nonzero eigenvalues come in pairs $\pm \lambda_n$ and the Dirac operator $\slashed{D}$ can be seen to have a symmetric spectrum. Although it may not be immediately obvious from looking at \eqref{Uchiral2}, when we make the summation explicit it becomes clear that the eigenvalues can be represented formally \citep{Tonggaugetheory} by the sequence $\lambda_1, -\lambda_1, \lambda_2, -\lambda_2, \lambda_3, -\lambda_3, \dots$, with each $\lambda_n$ possessing a corresponding $-\lambda_n$. Furthermore, the chiral nature of $\gamma_5$ means that it switches eigenfunctions with opposite eigenvalues $\gamma_5 \phi_n = \pm \phi_n$ \citep{Tonggaugetheory}. The nonzero $\phi_n$ then naturally pair up with eigenvalues of $\pm 1$ under the $\gamma_5$ operator, and the functional trace becomes

\begin{eqnarray}\label{FujikawaSumExample}
(\det e^{2i\gamma^5})^{-1} 
&=& \exp\left(-2i \left( \underbrace{0 + 0 + \dots + 0}_{\text{zero modes}} + \underbrace{1 + (-1) + 1 + (-1) + \dots}_{\text{paired modes}} \right) \right),
\end{eqnarray}

resembling a generalised alternating series \cite{Fujikawa:1980eg, Bertlmann:1996xk, Fujikawa:1999ku}.

Interestingly, for the case $\lambda_n = 0$ where we have the zero eigenvalues (i.e., zero modes), the operators $\slashed{D}$ and $\gamma_5$ can be simultaneously diagonalised since both $\phi_n$ and $\gamma^5 \phi_n$ have the same zero eigenvalue. This follows from the fact that $(\gamma_5)^2 = 1$, and the possible eigenvalues of $\gamma_5$ are $\pm 1$ \citep{Tonggaugetheory}. For these reasons, famously the index of the Dirac operator is written as $\text{index}(\slashed{D}) = n_+ - n_-$, with $n_+$ the number of zero modes with positive $\gamma_5$ eigenvalue and $n_-$ the number of zero modes with negative $\gamma_5$ eigenvalue \cite{Fujikawa:1979ay,Fujikawa:1999ku,Fujikawa:2004cx, Tonggaugetheory}. The zero modes are special \cite{Rylands2021ChiralAI}, since they are the only modes that contribute asymmetrically to the trace. Hence \citep{Fujikawa:1979ay}

\be
\int d^4x \, \sum \limits_{n=1}^{\infty} \phi^{\dagger}_n(x) \gamma_5 \phi_n(x) = \int d^4x \, \sum \limits_{n=1}^{\infty} \mid \phi_n \mid^2 h_n = \sum \limits_{n \in \, \text{zero modes}}^{\infty} h_n = n_+ - n_-,
\ee
where $h_n = \pm 1$ and
\be\label{IndexTheoremChiralAnomaly}
\text{index}(i\slashed{D}) = n_+ - n_- = \frac{1}{16\pi^2} \int d^4 x \, \epsilon^{\mu \nu \alpha \beta}F_{\mu \nu}F_{\alpha \beta},
\ee
which is the Atiyah-Singer index theorem \citep{AtiyahSinger1, Nielsen:1977aw,Tonggaugetheory}.

For the case $\lambda_n \neq 0$ the unregularised sum  clearly runs \textit{over all eigenvalues} of $\slashed{D}$. The Dirac operator $\slashed{D}$ and $\gamma_5$ now cannot be simultaneously diagonalised \citep{Bertlmann:1996xk}. Since for every eigenfunction $\phi_n$ with eigenvalue $\lambda_n$ there is another eigenfunction $\phi_{-n}$ with eigenvalue $-\lambda_n$, and since $\phi_n$ and $\phi_{-n}$ are mixed by $\gamma_5$, the Dirac operator $\slashed{D}$ and $\gamma_5$ do not commute. Thus, as a result, the spectral summation over the eigenmodes in \eqref{FujikawaSumExample} is \emph{conditionally convergent} \citep{Bertlmann:1996xk}. And it is a property of such sums that the order of summation affects the outcome of that summation. Heuristically, it is in this manner that the anomaly is ill-defined.

In the naive summation, one might expect the series \eqref{FujikawaSumExample} organised in periods of two $(1-1)+(1-1)+(1-1)+\dots$ to cancel to zero. However, the introduction of the smoothed cutoff $f\left(\slashed{D}^2/\Lambda^2\right)$ breaks the naive symmetry, leading to a net chiral \textit{asymmetry} \citep{Fujikawa:1980eg,Bertlmann:1996xk}. This is because the diagonalisation of $\slashed{D}$ that results in chiral asymmetry directly corresponds to a particular ordering of the summation \cite{Bertlmann:1996xk}, modifying the sum by damping the large eigenvalues (i.e., high energy modes) to make the series converge. Hence why, although the diagonalisation of $\slashed{D}$ reveals its spectral properties in terms of the pairings of eigenmodes, with exception of the zero modes \cite{Rylands2021ChiralAI}, upon the use of Fujikawa's regularisation the result is the imposition of gauge invariance in the trace \citep{Bertlmann:1996xk} but at the cost of the chiral (axial) symmetry becoming anomalous. Rather than the naive expectation that the trace would vanish $\Tr\left(\gamma_5\right)=0$, reflecting the symmetry in the eigenvalue contributions when the infinite series is organised in blocks of period two or period four $(1+1-1-1)+(1+1-1-1) + \dots =0$, instead we have $\Tr\left(\gamma_5\right) \neq 0$.


To understand more deeply the role of Fujikawa's regularisation, consider the infinite series $1+1-1-1+\dots$ which can be written as the following number theoretic sum\footnote{This series also has the intuitive representation as the sum $\sum \limits_{n=1}^{\infty} a_n$ with $a_n = \cos(n \pi/2) + \sin(n \pi/2)$.}\footnote{Thanks to Tony Padilla for discussions in relation to this series and its corresponding infinite sum.}

\be\label{FujikawaDivergentSum}
\sum \limits_{n=1}^{\infty} (-1)^{n(n-1)/2}.
\ee

Fujikawa's smooth cutoff $f\left(\slashed{D}^2/\Lambda^2\right)$ can be seen to effectively smooth out the sum of the eigenstates from $\slashed{D}$ in the sense defined in \citep{PadillaSmith1, Tao11}. In this way, Fujikawa's regularisation can be interpreted as an example of $\eta$ regularisation and its smoothed asymptotics \citep{PadillaSmith1,PadillaSmith2}. To see this, note that for Fujikawa's choice of regularisation the regularised infinite sum takes the form
\be\label{FujikawaDivergentSumReg1}
\sum \limits_{n=1}^{\infty} (-1)^{n(n-1)/2} e^{-n^2/\Lambda^2},
\ee
from which, using the method of smoothed asymptotics, we infer
\be\label{FujikawaDivergentSumReg2}
\lim \limits_{\Lambda \to \infty} \sum \limits_{n=1}^{\infty} (-1)^{n(n-1)/2} e^{-n^2/\Lambda^2} = 1.
\ee

This is a key number theoretic insight into Fujikawa's formulation of the path integral. When regularising the trace \eqref{Uchiral2}, what one is actually doing is weighting the infinite sum in the sense described in \cite{PadillaSmith1} and as first introduced from a number theoretic perspective in \cite{Tao11}. One finds that the contributions from the nonzero modes cancel pairwise in the smoothed asymptotics (owing to the pairing of eigenstates of opposite chirality), leaving behind the net contribution from the unpaired zero modes.

In terms of the regularised functional trace \eqref{Uchiral2FujiReg1} explicitly, let us demonstrate the example for $\alpha(x) = 1$ for concreteness. Suppose there are $k$ zero modes, and let us index the remaining paired modes by $\pm n$. Given the insertion of $f(\slashed D^2 / \Lambda^2)$, for a fixed choice of $\Lambda$ (i.e. before the limit is taken) one sees that the higher modes $\phi_n$ are indeed downweighted in the summation, with the regularised trace taking the form

\begin{eqnarray}
\lim_{\Lambda \rightarrow \infty} 
\exp\left(-2i \left( \underbrace{0 + \dots + 0}_{\text{$k$ zero modes}} + \underbrace{\frac{\lambda_{n}^2}{\Lambda^2}(1 - 1) + \frac{\lambda_{n+1}^2}{\Lambda^2}(1 - 1) + \dots}_{\text{paired modes}} \right) \right) = 1\, .
\end{eqnarray}

As the limit $\Lambda \rightarrow \infty$ is taken, the function $f(\slashed D^2 / \Lambda^2)$ converges pointwise to $1$ as expected. The overall effect of this regulation scheme is thus clearly equivalent to regularising the summation by applying a fixed ordering to its terms, and applying smooth asymptotics \citep{PadillaSmith1} to determine the result. In particular, the terms are ordered by increasing $|\lambda_n|$. 

\section{$\eta$ regularisation of the functional measure}
Mathematically, from the view of conventional measure theory (see for example \cite{rudin1966real,albeverio2008mathematical}), defining a measure on an infinite-dimensional space is highly non-trivial. Unlike in finite dimensions, where the Lebesgue measure provides a natural framework for integration, functional integration in infinite dimensions faces fundamental challenges. In particular, key properties of the Lebesgue measure---such as translation invariance and countable additivity---are often difficult to maintain in functional integration in QFT and related contexts.

For practical purposes, Lebesgue integrability is directly analogous to the absolute convergence of an infinite series: just as absolute convergence ensures the order of summation does not affect the sum, Lebesgue integrability ensures an integral is well-defined without requiring regularisation. However, as discussed in the previous section, the functional trace appearing in the chiral anomaly computation is only conditionally convergent. This means that its result depends on the order of summation, introducing ambiguities in defining the trace. The key mathematical insight is that this issue is directly analogous to the study of non-integrable functions in Lebesgue theory, where improper integrals must be carefully regularised to produce finite results. Thus, just as conditionally convergent series require summation rules, the functional trace in the chiral anomaly calculation requires a well-defined regularisation scheme to control divergences.

Unlike conventional measures, the path integral measure does not necessarily satisfy all the axioms of the Lebesgue measure in a strict mathematical sense. As a result, alternative formulations such as the Wiener measure (e.g., for stochastic processes) or the Schwinger proper-time representation are sometimes employed in special cases. However, these approaches do not fully resolve the challenge of defining a general, well-behaved path integral measure in QFT \cite{Fujikawa:2004cx,albeverio2008mathematical}. Furthermore, if we adopt the viewpoint that the path integral measure lacks a natural Lebesgue definition due to these issues with infinite-dimensional integration, then introducing a regularisation scheme serves a role analogous to ensuring Lebesgue integrability in finite-dimensional analysis. Specifically, instead of considering an unregularised functional measure, we can define a regularised measure incorporating a smoothing operator that ensures well-defined convergence and controls divergences systematically.

More formally, this analogy suggests defining an operator-valued regularisation scheme within the definition of the functional measure. That is, we are motivated to define a regulating function $\eta\left(\Ocal \right)$ which is then applied to the spectrum of the relevant operator $\Ocal$ (e.g., the squared Dirac operator $\slashed{D}^2$) in order to dampen high-energy modes and restore well-defined summation properties. This formulation naturally leads to an improved understanding of the role of smoothing in the functional determinant, linking the regularised measure to the trace-class properties of the relevant operators. 

The regularised functional measure is defined as

\be\label{RegularisedMeasureGen}
\Dcal \Phi \rightarrow \Dcal \Phi |_{\text{Regularised}} = \Dcal(\Rcal_{\eta} \cdot \Phi),
\ee
with $\Rcal_\eta$ the regulating operator. This regulating operator is then defined in terms of a smooth cut-off function $\eta(\Ocal)$ motivated by the $\eta$ regularisation scheme \cite{PadillaSmith1, PadillaSmith2} such that
\be\label{RegulatingOp}
\Rcal_{\eta} = \eta_{\Lambda}\left(\mathcal{O}\right),
\ee
where $\eta_{\Lambda}\left(\Ocal \right)$ is a real analytic function $\eta : [0, \infty) \to [0, 1]$. More general than $\eta$ regularisation for loop integrals, the smooth regulating function $\eta_{\Lambda}\left(\mathcal{O}\right)$ accepts an operator-valued input $\mathcal{O}$, where $\Ocal$ is assumed to be a positive-definite (non-defective) linear operator on a separable Hilbert space $\mathcal{H}$ and has a discrete set of eigenvalues $\lambda_n$. This means that $\Ocal$ is assumed to admit a spectral decomposition of the form
\be
\Ocal = \sum_n \,\lambda_n\,|\psi_n\rangle\langle\psi_n|,
\ee
where $\{|\psi_n\rangle\}$ is a complete orthonormal basis of eigenfunctions.  In other words, for each $n$ it follows $\Ocal\,|\psi_n\rangle = \lambda_n\,|\psi_n\rangle$ and $\langle\psi_m|\psi_n\rangle = \delta_{mn}$. From this expansion, it also follows that for any orthonormal basis $\{|e_k\rangle\}$ of $\mathcal{H}$,
\be
\Tr(\Ocal) = \sum_{k}\,\langle e_k|\mathcal{O}|e_k\rangle =  \sum_{k}\,\sum_{n}\,\lambda_n\,\langle e_k|\psi_n\rangle\langle\psi_n|e_k\rangle.
\ee

However, since $\{|\psi_n\rangle\}$ is a complete orthonormal basis, we find

\be
  \sum_{k}\,\langle e_k|\psi_n\rangle\langle\psi_n|e_k\rangle
  \;=\;
  1
  \quad\text{(resolution of the identity)}.
\ee

Hence, as a standard result, we arrive at the usual functional trace

\be\label{OperatorTrace}
  \mathrm{Tr}(\mathcal{O})
  \;=\;
  \sum_n\,\lambda_n \langle\psi_n| \psi_n \rangle,
\ee
which is the standard statement that the trace of the operator $\Ocal$ is the sum of its eigenvalues. From Lidskii's theorem (i.e., spectral theorem), for the sum to be well-defined $\Ocal$ must be at least a trace-class operator \cite{conway2000course,reed1980methods} such that $\sum \limits_n |\lambda_n| < \infty$. Many physically relevant operators fail to satisfy this condition, particularly unbounded operators such as differential operators (e.g., the Laplacian $\Delta$, the Dirac operator $\slashed{D}$) and the chirality operator $\gamma_5$. In these cases, the naive sum of eigenvalues diverges, much like the ill-defined infinite sum encountered in Fujikawa's computation of the chiral anomaly.

The notion of an operator-valued smooth regulator \eqref{RegulatingOp} therefore directly generalises Fujikawa's regularisation, since we define $\eta_{\Lambda}\left(\mathcal{O}\right)$ in terms of the eigenbasis of $\Ocal$. This is directly analogous to the procedure when defining the unregularised Fujikawa measure \eqref{Pathintegralmeasure} as reviewed in earlier, and it will even allow us to explicitly capture Fujikawa's regularisation scheme. This is because, in an analogous way to Fujikawa's regularisation, the domain of $\eta(x)$ corresponds to the spectrum of the operator $\Ocal$. For operators like $ \slashed{D}^2$, the eigenvalues are non-negative. The codomain ensures that $\eta(x)$ serves as a smoothly regulating weight satisfying the properties that $\eta(0) = 1$ for small $x \rightarrow 0$ (coincidentally, $ \lim_{\Lambda \to \infty}$) and $\eta(x) \to 0$ rapidly as $x \to \infty$ (coincidentally, $\lim_{\Lambda \to 0}$).

To make sense of the proposed generalised framework, recall the definition of the measure \eqref{Pathintegralmeasure}. The product signifies that each mode $c_n$ contributes independently to the integration. For the general case of the regulated measure \eqref{RegularisedMeasureGen}, the strategy is similar. Let the operator $\Ocal$ be interpreted according to its operation on $\Phi$ in \eqref{RegularisedMeasureGen}, and let $\Tr [\Ocal]_{\Dcal(\eta(\Ocal) \cdot \Phi)}$ denote the trace of $\Ocal$ with respect to the regularised measure $\Dcal(\eta(\Ocal) \cdot \Phi)$. Then, for a complex-valued field $\phi$, we may express the decomposition of the regularised measure as
\be\label{RegularisedMeasureDecomp}
D(\eta(\Ocal) \cdot \phi) = \bigwedge \limits_{n=1}^{\infty} d\phi^{\prime}_n = \bigwedge \limits_{n=1}^{\infty} d(\lambda_n \phi_n),
\ee
where $\{\phi^{\prime}_n\}_{n=1}^{\infty}$ represents any standard orthonormal basis of the Hilbert space associated with the functional space of the field $\phi^{\prime}$. Here $\lambda_n$ are the eigenvalues of the regulator accepting operator valued inputs and $\phi_n$ are the corresponding eigenfunctions. Note that, instead of the naive product appearing previously \eqref{Pathintegralmeasure}, the wedge product is used in the generalised case since we are dealing with functional determinants and measures in an infinite-dimensional space \cite{reed1980methods}. 

Furthermore, we denote $\bigwedge \limits_{n=1}^\infty d\phi^{\prime}_n$ to describe the infinitesimal volume form constructed by taking the ``products" of all infinitesimal differentials $d\phi^{\prime}_n$ associated with the orthonormal basis components. We therefore consider the regularised measure to be over \textit{all degrees of freedom} in the Hilbert space, except for the case $\lambda_j = 0$ implying that the zero modes of $\eta(\Ocal)$ are omitted. This formulation clearly aligns with standard approaches in functional analysis to defining orthonormal bases, where, it will be remembered, the basis functions $\{\phi_n\}$ diagonalise an operator.

Generally speaking, defining $\Rcal_{\eta} = \eta_{\Lambda}(\Ocal)$, we treat $\Rcal_{\eta}$ as a linear regulating operator that acts on the entire functional field space. We assume it is a self-adjoint operator and has a discrete spectral decomposition in the Hilbert space of field configurations. This construction also means that the action of $\eta_{\Lambda}\left(\Ocal \right)$ on field configurations $\Phi$ is determined by expressing $\Phi$ in the eigenbasis of $\Ocal$. Since linear operators preserve eigenbasis action \cite{Feynman1965QuantumMA,Ramond:1981pw,Fujikawa:2004cx} for any eigenstate $|n\rangle$ of $\Ocal$ with eigenvalue $\lambda_n$ such that $\Ocal |n\rangle = \lambda_n |n\rangle$, the same properties are satisfied in the regularised case. This follows from the fact that the regulator $\eta_{\Lambda}\left(\Ocal \right)$, by definition, operates by transforming the eigenvalues of $\Ocal$; therefore, applying $\eta_{\Lambda}\left(\Ocal \right)$ to $|n\rangle$ simply modifies the eigenvalue.

For these reasons, analogous to Fujikawa's framework, given the regulator $\eta_{\Lambda}\left(\Ocal \right)$ is a real-valued function that acts on the operator $\Ocal$, and given the eigenstates $|n \rangle$ and the corresponding eigenvalues $\lambda_n$ of $\Ocal$, the action of the regulator on the eigenstates is simply defined by

\be\label{Regularisedeigenvalues}
\eta_{\Lambda}\left(\Ocal\right) \cdot |n\rangle = \eta_{\Lambda}\left(\lambda_n \right) |n \rangle.
\ee

Since $\eta_{\Lambda}(\Ocal)$ is defined consistently in terms of the spectral analysis, it is easy to see that this argument can be extended for the action of the regulator on any general state $|\Phi \rangle$ by linear combination of its action on the basis states

\be
|\Phi\rangle = \sum_n c_n |n\rangle \quad \rightarrow \quad \eta(\Ocal)|\Phi\rangle = \sum_n c_n \eta_{\Lambda}(\lambda_n) |n\rangle,
\ee

assuming that this holds for any $|\Phi\rangle$ that belongs to the domain of $\eta\left(\Ocal \right)$, ensuring the regulator is well-defined on the chosen Hilbert space. Since $\eta\left(\Ocal \right)$ suppresses high-energy modes, the resulting sum is absolutely convergent, thereby recovering the smoothed sum \eqref{FujikawaDivergentSumReg1} for the choice of Fujikawa's regularisation. 

\subsection{Recovering the chiral anomaly}
To check that our generalised framework is consistent with Fujikawa's path integral formalism, in this section we show that we can recover the correct expression for the chiral anomaly as reviewed in Section \ref{Fujikawa} and Section \ref{EigensumSeries}. As discussed, there are two complementary views of the chiral anomaly. In the standard Fujikawa approach, the regulator plays an essential role in damping UV divergences and breaking the spectral symmetry of the naive trace $\Tr(\gamma_5)$, thereby revealing the anomaly. In contrast, the index theorem approach shows that the anomaly localizes on the zero modes, where the regulator becomes trivial and the trace reduces to the index of the Dirac operator. These perspectives are reconciled by the fact that the anomaly is a topological quantity, independent of the regulator's detailed form. This picture naturally emerges in the case of the regularised measure.

For simplicity, and to directly access the structure of the anomaly, consider the density over field configurations $\Phi$ for the Euclidean path integral

\be\label{NonRegMeasureProbDensity}
P = \Dcal \Phi \exp\left(S[\Phi]\right).
\ee

Since only the matter content contributes to the anomaly, this formal representation shall be sufficient. Once again assuming a massless theory, the regularised density takes the form

\be\label{RegMeasureProbDensity}
P_{\Rcal_{\eta}} = \Dcal \left(\Rcal_\eta \cdot \Phi \right) \exp\left(S[\Phi]\right).
\ee

Let $U(x)$ be a local symmetry as defined in Section \ref{Fujikawa}. We write $U(\Phi)$ to denote the field $\Phi$ transformed pointwise by the local operator $U(x)$ such that $[U(\Phi)](x):= U(x)\Phi(x)$. It acts as a transformation on $\Phi$ in both the regularised measure and the classical action such that

\be\label{RegMeasureClassicalSym}
\Dcal \left(R_\eta \cdot \Phi\right) \exp\left(S[\Phi]\right) = \Dcal \left(R_\eta \cdot U(\Phi)\right) \exp\left(S[U(\Phi)]\right).
\ee

If $U(x)$ is linear, the fields and the action transform as $\Phi \to \Phi^{\prime} = U(x)\Phi(x)$ and $S[\Phi] \to S[U(\Phi)]$, respectively. Clearly, in the non-regularised case, \eqref{NonRegMeasureProbDensity} transforms as $P \to P^{\prime} = \Dcal\left(U \cdot \Phi \right)\exp\left(S[U(\Phi)]\right)$. For the regularised measure, it is similarly straightforward to show that the regularised density \eqref{RegMeasureProbDensity} transforms as
\begin{eqnarray}\label{RegMeasureProbDensityTrans1}
P_{\Rcal_{\eta}}^{\prime} &=& \mathcal D \left(\Rcal_\eta \cdot U(\Phi)\right) \exp\left(S[U(\Phi)]\right) \\
&=& \Dcal \left(\Rcal_\eta \cdot \Phi\right) \det \left(J\right) \exp\left(S[U(\Phi)]\right),
\end{eqnarray}
where $\left(J\right) = \Rcal_\eta U \Rcal_\eta^{-1}$ is the transformation operator, whose determinant captures the change in the measure. The transformed action $S[U(\Phi)] = S[\Phi] + \Delta S$ with $\Delta S = S[U(\Phi)] - S[\Phi]$ explicitly contains the classical symmetry breaking. Therefore, we see that the regularised action is modified by two distinct contributions
\be\label{RegMeasureAnomaly1}
P_{\Rcal_{\eta}}^{\prime} = \Dcal \left(\Rcal_\eta \cdot \Phi\right)\exp\left(S[\Phi]\right)\exp\left(\Delta S + \ln \det \left(J\right)\right),
\ee
with $\Delta S$ the change to the action functional and $\ln \det \left(J\right)$ is the Jacobian dependent term capturing the quantum anomaly resulting from the regularised measure. Rearranging terms in \eqref{RegMeasureAnomaly1}, we can rewrite the regularised density so that
\be
P_{\Rcal_{\eta}}^{\prime} = P_{\Rcal_{\eta}} \exp\left(\Delta S + \ln \det (J)\right),
\ee
where $P_{\Rcal_{\eta}}$ is defined as in \eqref{RegMeasureProbDensity}. Dividing both sides by $P_{\Rcal_{\eta}}$ and taking the logarithm, we obtain the logarithmic variation of the regularised probability density

\be
\ln \left(\frac{P^{\prime}_{\Rcal_\eta}}{P_{\Rcal_\eta}} \right) = \Delta S + \ln \det (J).
\ee

Defining the variation $\Delta \ln P_{\Rcal_{\eta}}^{\prime} := \ln \left(\frac{P^{\prime}_{\Rcal_\eta}}{P_{\Rcal_\eta}} \right)$ so that
\be\label{RegMeasureAnomaly2}
\Delta \ln P_{\Rcal_{\eta}}^{\prime} = \Delta S + \ln \det (J),
\ee
the right-hand side can be seen to contain the sum of the classical symmetry breaking and the quantum symmetry breaking. Let $U(x)$ be small such that $U=\exp\left(i\alpha(x) T \right)$ for infinitesimal $\alpha(x)$, which is a function of position. As the trace is over field configurations, we have
\be\label{RegMeasureAnomaly3}
\Delta \ln P_{\Rcal_{\eta}}^{\prime} = \Delta S + i\int d^4 x \, \alpha(x) \Tr (\Rcal_{\eta}^{\dagger} \Rcal_{\eta} T),
\ee
via the standard identity $\det(e^{i\alpha(x) T}) = e^{i\alpha(x) \Tr(T)}$ giving the operator-theoretic form of the regularised trace as a local density integrated over spacetime. 

The trace $\Tr (\Rcal_{\eta}^{\dagger} \Rcal_{\eta} T)$ is over internal (spinor and gauge) indices. Recalling $\Rcal_{\eta} = \eta_{\Lambda}(\Ocal)$, we introduce a regulator for $\Rcal$ and $\Rcal^{\dagger}$ such that $\Rcal = \eta(\Ocal)$ and $\Rcal^{\dagger} = \tilde{\eta}(\Ocal)$. Following Fujikawa and Suzuki \citep{Fujikawa:2004cx}, the key observation is that from \eqref{RegMeasureAnomaly3} we may thus proceed either by analysing $\psi$ and $\bar{\psi}$ separately, or with the covariant calculation, which, at the end of the day, because of gauge covariance, amounts to the two regulators factorising into a single regulator function. In the covariant approach, this effectively amounts to a single insertion $\Rcal_{\eta} \gamma_5$ upon specialising to the chiral transformation for $T = \gamma_5$.\footnote{Otherwise the regularised measure would suffer mathematical inconsistency and inconsistency with Fujikawa’s original derivation of the chiral anomaly.}  A natural choice for the smooth cut-off is of course Fujikawa's regulator $\eta(\Ocal) = e^{-\slashed{D}^2/\Lambda^2}$. Making this choice, it is straightforward to see that the regularised measure yields the standard expression for the chiral anomaly, as reviewed in Section \ref{Fujikawa}. If, however, we instead follow a similar strategy as in \citep{PadillaSmith1}, keeping the precise form of $\eta$ general, it follows that we obtain for the regularised trace

\be\label{RegMeasureAnomaly4}
\Tr[\Rcal_{\eta} \gamma_5] = i \int d^4x \, \alpha(x)\Tr[\gamma_5 \langle x | \eta \left(\frac{\slashed{D}^2}{\Lambda^2}\right) |x\rangle].
\ee

To recover the chiral anomaly for general $\eta$, we recall the results from Section \ref{EigensumSeries}. As discussed, by inserting the smooth regulator $\eta(\slashed{D}^2/\Lambda^2)$ we recognise the ill-defined trace can be written as
\be
\Tr\left( \gamma_5 \eta \left(\frac{\slashed{D}^2}{\Lambda^2}\right) \right)
= \sum_n \langle \phi_n | \gamma_5 \eta \left( \frac{\lambda_n^2}{\Lambda^2} \right) | \phi_n \rangle,
\ee
which weights the eigenvalue sum and breaks the symmetric spectrum. As $\Lambda \to \infty$, the regulator becomes $1$ on the zero modes and $0$ on the high eigenmodes, effectively extracting the contribution of the zero modes. Furthermore, when $\Lambda \to \infty$, the regularised trace in the limit projects onto the kernel of $ \slashed{D}^2$ because we demand $\eta(0) = 1$ ensuring the zero modes are included fully, and $\eta(\lambda_n^2/\Lambda^2) \to 0$ for $ \lambda_n \ne 0$, so nonzero modes are suppressed. Thus, the trace becomes

\be
\sum_{\lambda_n = 0} \langle \phi_n | \gamma_5 | \phi_n \rangle = n_+ - n_- = \text{index}(\slashed{D}),
\ee
which recovers the topological identity \eqref{IndexTheoremChiralAnomaly} for the chiral anomaly. The same follows for $\Rcal^{\dagger} \gamma_5$. Although the chiral anomaly is ultimately captured by the index of the Dirac operator, and hence by the contribution of the zero modes, this index arises through a regularised summation over all eigenmodes of the Dirac operator. In this way, the anomaly is ``spread" across the spectrum in the regulated trace, even though the unregulated sum would vanish. Only in the $ \Lambda \to \infty$ limit does the regulator act as a projection onto zero modes, at which point the anomaly becomes a topological invariant.

In conclusion, while the action $S[\Phi]$ may exhibit a classical symmetry, this symmetry is not necessarily preserved in the regularised measure $\Dcal\left(\Rcal_\eta \cdot \Phi\right)$. As demonstrated, our generalised regularisation framework not only reproduces the known expression for the chiral anomaly, but also provides a clear connection between the formal structure of the path integral, the spectral nature of the Dirac operator, and the topological content captured by the index theorem.

\subsection{Studying the effects of the regularised measure}\label{SecIota}
In $\eta$ regularisation for loop integrals \citep{PadillaSmith1, PadillaSmith2}, we note that a key mechanism which ensures all divergences are completely killed comes from vanishing Mellin transforms of the smooth cut-off. In this section, we explore if there exists an analogous mechanism in the regularised measure \eqref{RegularisedMeasureGen}.

To do this, our strategy is to probe the measure's transformation under a change to the regularisation scale $\Lambda$. We therefore start by varying the cut-off scale $\Lambda$, which will enable us to study the trace of the derivative of the regulator kernel. After some careful manipulations, we then systematically isolate the high-energy eigenvalue contributions and examine the behaviour of the regularised measure, particular as a means to probe the analytic structure of the cut-off and whether any non-trivial cancellations occur. This approach is closely related to Fujikawa's \citep{Fujikawa:2004cx}, and it is expected that the physical result of the chiral anomaly is independent of the regularised measure.

Consider again the functional measure in the regularised density \eqref{RegMeasureProbDensity} for the case of a fermionic field $\psi$. Taylor expanding the regularising operator $\Rcal_{\eta} (\Lambda+\epsilon) = \Rcal_{\eta} + \epsilon \frac{\mathrm{d}\Rcal_\eta}{\mathrm{d}\Lambda} + \mathcal{O}(\epsilon^2)$ in small $\epsilon$, the regularised measure can be written as

\begin{eqnarray}\label{ExpandedRegOp}
P^{[\psi]}_{\Rcal_\eta} (\Lambda + \epsilon) &=& \Dcal \left(\Rcal_{\eta} \left(1 + \epsilon \Rcal^{-1}_{\eta} \frac{\mathrm{d} \Rcal_\eta}{\mathrm{d}\Lambda}\right) \cdot \psi \right) \\ \nonumber
&=& \Dcal\left(\Rcal_{\eta} \cdot \psi \right) \det \left(1 + \epsilon \Rcal_\eta^{-1} \frac{\mathrm{d} \Rcal_\eta}{\mathrm{d}\Lambda}\right)^{-1}.
\end{eqnarray}

For any change of variables the measure rescaling depends on the Jacobian determinant. In this case, the determinant captures the infinitesimal change in the regularised measure due to the variation of the cut-off scale. To write this in a more suggestive form, in the second line of \eqref{ExpandedRegOp} we recognise $\Dcal\left(\Rcal_{\eta} \cdot \psi \right)$ as $P_{\Rcal_\eta}$, since we are only concerned here with the measure part of \eqref{RegMeasureProbDensity}. Assuming $\epsilon \Rcal_\eta^{-1} \frac{\mathrm{d} \Rcal_\eta}{\mathrm{d}\Lambda} << 1$, which holds if $\Rcal_\eta$ varies smoothly with $\Lambda$, we can use the standard expansion $\det (1 + A) = \exp(\Tr[A])$ to rewrite the determinant. Taking these steps we get

\be
P^{[\psi]}_{\Rcal_\eta} (\Lambda + \epsilon) = P^{[\psi]}_{\Rcal_\eta} \exp\left(-\Tr\left[\epsilon \Rcal_\eta^{-1} \frac{\mathrm{d} \Rcal_\eta}{\mathrm{d}\Lambda}\right]\right),
\ee
where we have included a minus sign in the exponential coming from the fact that the measure transformation includes a fermionic field $\psi$, and determinants in fermionic functional space are inverted. Since we are interested in studying the effects of the regularisation as the regulator is lifted, and since the variation of the regulator captures the change in the eigenvalue spectrum, we can rewrite the determinant purely in terms of $[d\Rcal_\eta / d\Lambda]$. Then, taking the logarithm of both sides and rearranging terms, we obtain

\be
\frac{\ln P^{[\psi]}_{\Rcal_\eta} (\Lambda + \epsilon) - \ln P^{\psi}_{\Rcal_\eta}}{\epsilon} = -\Tr\left[\frac{\mathrm{d}\Rcal_\eta}{\mathrm{d}\Lambda}\right].
\ee

It is important to emphasise that, on the left-hand side, we are varying only the measure part. On the right-hand side, $\Rcal_\eta$ is an operator acting in field space, so the trace must be defined with respect to an appropriate basis. 

The next step is to observe that for small $\epsilon$ the left-hand side may be approximated by a first-order finite difference. We therefore arrive at the relation describing the change in the measure due to incremental change in the cut-off scale\footnote{Note, one can perform the same calculation for the case of a bosonic field, and a similar result follows analogously.}

\be\label{DensitytransfFermion}
\frac{d}{d\Lambda} \ln P^{[\psi]}_{\Rcal_\eta} = -\Tr\left[\frac{\mathrm{d} \Rcal_\eta}{\mathrm{d}\Lambda}\right].
\ee

Using the chain rule for logarithmic differentiation, we express the left-hand side as a ratio of logarithms, which aligns with structures commonly found in renormalisation group equations \citep{Salmhofer2007RenormalizationAI}. As a consequence of this rewriting we pick up a factor of $\Lambda$ on the right-hand side. Since the fermionic path integral measure involves integration over both $\psi, \bar{\psi}$ , the total contribution doubles. Considering this, we arrive at the relation defined in terms the function $\iota_E(\Lambda)$

\be\label{iotadefeucl}
\iota_E(\Lambda) \, := \frac{\mathrm d (\ln P^{[\psi, \bar{\psi}]}_{\Rcal_\eta})}{\mathrm d (-\ln \Lambda)} = -2 \Lambda \Tr \left[{\frac{\mathrm d \Rcal_\eta^2}{\mathrm d \Lambda}}\right].
\ee

To ensure consistency with the fermionic path integral structure, following a similar strategy as discussed in the previous section, we write $\Rcal_\eta^2 = \Rcal_\eta \Rcal_\eta^\dagger$ and can thus proceed with the covariant calculation. We then set $\Rcal_{\eta} = \eta\left(\Ocal \right) = \eta \left(\slashed{D}^2 / \Lambda^2 \right)$ in \eqref{iotadefeucl}, keeping the explicit form of the regulator general. We also note that \eqref{iotadefeucl} is beginning to look like a flow equation for the path integral measure under infinitesimal scaling of the cut-off. At this stage, the quantity $\iota_E(\Lambda)$ measures how sensitive the regularised measure is to changes in the cut-off.

Before calculating the trace, we use the chain rule again to rewrite $\iota_E(\Lambda)$ as

\be
\iota_E(\Lambda) = \Lambda \, \mathrm{Tr} \left[ \frac{\mathrm{d} \left( \eta \left( \frac{\slashed{D}^2}{\Lambda^2} \right) \right)}{\mathrm{d} \Lambda} \right] = -\frac{2}{\Lambda^2} \mathrm{Tr} \left[ \eta' \left( \frac{\slashed{D}^2}{\Lambda^2} \right) \slashed{D}^2 \right].
\ee 

Since $\eta(x)$ acts on the eigenvalues of $\slashed{D}^2$, the trace naturally transforms as $\eta' \left( \frac{\slashed{D}^2}{\Lambda^2} \right) \slashed{D}^2$, ensuring that only eigenvalue contributions appear in the result. 

Next, using Fujikawa's plane wave analysis \citep{Fujikawa:2004cx}, we may write

\be
\iota_E(\Lambda) = -\frac{2}{\Lambda^2} \int \mathrm{d}^4x \langle x | \Tr \left[ \eta' \left( \frac{\slashed{D}^2}{\Lambda^2} \right) \slashed{D}^2 \right] | x \rangle
\ee
from which it follows
\[
= -\frac{2}{\Lambda^2} \int \mathrm{d}^4x \int \frac{\mathrm{d}^4k}{(2\pi)^4} \langle x | \Tr \left[ \eta' \left( \frac{\slashed{D}^2}{\Lambda^2} \right) \slashed{D}^2 \right] | k \rangle \langle k | x \rangle
\]

\[ = -\frac{2}{\Lambda^2} \int \mathrm{d}^4x \int \frac{\mathrm{d}^4k}{(2\pi)^4} \langle k | x \rangle \langle x | \Tr \left[ \eta' \left( \frac{\slashed{D}^2}{\Lambda^2} \right) \slashed{D}^2 \right] |k \rangle
\]

\be\label{iotaplanewave}
= -\frac{2}{\Lambda^2} \int \mathrm{d}^4x \int \frac{\mathrm{d}^4k}{(2\pi)^4} \Tr \left[ \eta' \left( \frac{\slashed{D}_k^2}{\Lambda^2} \right) \slashed{D}_k^2 \right] \cdot 1_x,
\ee
where $1_x$ is the resolution of identity in $x$-space.

In the transition from the second-last to last line in \eqref{iotaplanewave}, any differential operators $\partial_\mu$ in the trace over fermion indices are replaced with $\partial_\mu + ik_\mu$, following plane-wave decomposition rules \citep{Fujikawa:2004cx}. This substitution is valid for local operators, ensuring the Dirac operator in momentum space takes the form $\slashed{D}_k = \gamma^\mu (D_\mu + ik_\mu)$, which is standard in Fujikawa's approach.\footnote{The procedure of converting operators in this way is well-defined for local operators: express an operator as a sum of products of operators which are either differential (i.e. diagonal in $k_\mu$ basis) or non-differential (i.e. diagonal in $x^\mu$ basis). Then replace $\partial_\mu \mapsto \partial_\mu + ik_\mu$. In flat spacetime, a single coordinate patch covers the whole space, so this representation holds for all (flat) spacetime.} This is the reason for the subscript in $\slashed D_k$. For concreteness, $k_\mu$ act like constants, not differential operators. They are merely the variables of integration referred to by the momentum integration $\mathrm d^4 k$.

Before proceeding, we may simplify \eqref{iotaplanewave} further by using change of variables $z = \slashed{D}_k^2 / \Lambda^2$. Doing so gives

\be\label{iotaplanewave2}
\iota_E (\Lambda) = -\frac{2}{\Lambda^2} \int \mathrm{d}^4x \int \frac{\mathrm{d}^4k}{(2\pi)^4} \Tr \left[ \eta' \left( z \right) \Lambda^2 z \right] \cdot 1_x.
\ee

Using the identity $\slashed D^2 = D_\mu D^\mu + F$ and the gauge covariant decomposition $F = 1/4 [\gamma^\mu , \gamma^\nu] [D_\mu, D_\nu]$, we can expand the derivative operator such that

\be
z = z_0 + \Delta, \, \text{with} \, z_0 = -k^2/\Lambda^2, \, \text{and} \, \Delta = \frac{2ik \cdot D + D^2 + F}{\Lambda^2}.
\ee

Assuming $\Delta$ is small, since the quadratic momentum term $-k^2/\Lambda^2$ dominates the argument of $\eta^{\prime}$ in the UV region, where $|k|$ is large, the rest of the terms can be seen to either scale sublinearly or are independent of $k$. Since $\eta$ is smooth and its derivatives are smooth, we can therefore expand $\eta^{\prime}(z)$ around $-k^2$ giving

\be
\eta'\Bigl(z_0+\Delta\Bigr)
=\eta'\Bigl(z_0\Bigr)
+ \eta''\Bigl(z_0\Bigr)\,\Delta
+ \frac{1}{2}\eta'''\Bigl(z_0\Bigr)\,\Delta^2
+\cdots\,.
\ee

This is practically analogous to the heat kernel expansion \citep{Ball:1988xg}. Putting everything together, the product in the integrand \eqref{iotaplanewave2} takes the form

\be
\eta'\Bigl(z_0+\Delta\Bigr)\,(z_0 + \Delta)
= \Biggl\{
\eta'\Bigl(z_0\Bigr)(z_0+\Delta)
+\eta''\Bigl(z_0\Bigr)(z_0+\Delta)\Delta
+\cdots
\Biggr\}\,.
\ee

At leading order, the term $\eta'(z_0)\,z_0$ corresponds to free propagation and is independent of the external fields. The next non–vanishing contributions come from terms linear or quadratic in $\Delta$ that contain factors like $D$ and $F$. Ultimately, to extract the chiral anomaly one requires terms quadratic in $F_{\mu \nu}$, which appear in the expansion of $\Delta$ and with $\Delta^2$ also containing terms quadratic in $D$. However, since the goal is to study how the measure depends on the regulator---to determine whether the regulator causes any unexpected contributions or cancellations---it is sufficient to focus only on the linear terms in $\Delta$. Thus, we obtain

\be\label{iotaplanewave2}
\iota_E (\Lambda) = -2 \int \mathrm{d}^4x \int \frac{\mathrm{d}^4k}{(2\pi)^4} \Tr \left[ \eta'(z_0)z_0 + \left[\eta'(z_0) + \eta''(z_0)z_0\right]\Delta
+ \mathcal{O}(\Delta^2) \right] \cdot 1_x.
\ee

Following the procedure in \citep{Fujikawa:2004cx}, the next step is to exploit the symmetry of the momentum integration. We apply standard symmetric integration identities to replace products of $k_\mu \dots k_\nu$ with products of $k^2 = k_\mu k^\mu g_{\mu \nu}$. Odd powers of $k^{\mu}$ vanish, meaning $\int \mathrm{d}^4k/(2\pi)^4 \, k^\mu f(k^2) = 0$. Accordingly, the momentum integrals in \eqref{iotaplanewave2} reduce to integrals of the general form

\be
I(n,m) = \int \frac{d^4k}{(2\pi)^4}\,\eta^{(n)}\!\left(-\frac{k^2}{\Lambda^2}\right)(-\frac{k^2}{\Lambda^2})^m,
\ee
which appears as a Mellin-like structure \citep{PadillaSmith1, PadillaSmith2}. As it turns out, this will be seen to not be a coincidence. Moreover, to extract the $\Lambda$ dependence from \eqref{iotaplanewave2}, we perform a change of variables in the $k$-integration. Let $t = k^2/\Lambda^2$ so that $k = \sqrt{t}\Lambda$ and $dk = dt \Lambda/(2\sqrt{t})$. Then, for the leading term, we find
\be
\int \frac{d^4k}{(2\pi)^4} \Tr\left[z_0 \eta^{\prime}(z_0)\right] = -\frac{\Lambda^4}{8\pi^2} \, \mathcal{M}_1 \{\eta \} \Tr[1],
\ee
where we've used integration by parts and where $\mathcal{M}_1\{\eta\}$ is the Mellin transform of the regulator \citep{PadillaSmith1}. For the next subleading term we get

\be
\int \frac{d^4k}{(2\pi)^4} \Tr\left[\eta^{\prime}(z_0) \Delta\right] = \frac{\Lambda^2}{16\pi^2} \mathcal{M}_0 \{\eta \} \Tr[F].
\ee

Altogether we find

\be\label{IotaFinalMellin}
\iota_E (\Lambda) = \frac{1}{4\pi^2} \int \mathrm{d}^4x \left( \Lambda^4 \mathcal{M}_1 \{\eta \}\Tr[1] - \frac{\Lambda^2}{2} \mathcal{M}_0 \{\eta \} \Tr[F] + \dots \right).
\ee

The leading terms in the limit \(\Lambda \to \infty\) (as the regulator is lifted) are those proportional to \(\Lambda^4\) and \(\Lambda^2\). These terms are divergent and depend explicitly on the regulator \(\eta\). We notice that the presence of the Mellin transforms resembles a structure similar to that found in \citep{PadillaSmith1} for which enhanced regulators ensured the divergences were completely killed. This observation serves as a nice motivation to draw deeper connection between different applications of $\eta$ regularisation. Since we want $\iota_E (\Lambda) = 0$ to vanish, if we choose $\eta$ to be enhanced for each respective Mellin integral, the overall integrand in \eqref{IotaFinalMellin} can be made to vanish.\footnote{In fact, what is likely required is the choice of a stricter class of ``super-enhanced" regulators. The development of a theory of ``super-enhanced" regulators extends beyond the remit of this work.}

\section{Bridging regularisation schemes: a unifying view}
It was observed in \citep{Ball:1988xg} that generalised Schwinger proper time integrals provide an elegant framework for calculating anomalies. In \citep{Umezawa:1989ns}, Umezawa showed that Fujikawa's regularisation is closely related to the Schwinger proper time approach in the coordinate representation. Most recently, in the context of perturbative loop integrals, it was shown \citep{PadillaSmith2} that the large class of $\eta$ regulators also correctly capture the chiral anomaly for the two dimensional Schwinger model. This general regularisation scheme is also closely connected to the generalised Schwinger proper time approach \citep{PadillaSmith1}.

The purpose of the present section is to understand these distinct schemes from a unified viewpoint, particularly in how they relate to our generalised framework for the regularised path integral measure. For this purpose, the generalised Schwinger proper time framework is used as an important conceptual bridge. 

We work in the Schwinger model, which is two dimensional Quantum Electrodynamics (QED$_2$) with massless Dirac fermions interacting with an abelian gauge field. The quantisation of the fermion field $\psi(x)$ follows from the Lagrangian density

\be\label{SchwingerLagragian}
\mathcal{L} = \bar{\psi}(i\slashed{D})\psi - \frac{1}{4}F_{\mu \nu} F^{\mu \nu}.
\ee

The Euclidean action is

\be\label{Schwingeraction}
S = \int d^2x \left[ \bar{\psi} (i \slashed{D}) \psi - \frac{1}{4} F_{\mu\nu} F^{\mu\nu} \right],
\ee
where $\bar{\psi} = \psi^\dagger \gamma^0$ is the Dirac adjoint spinor, $\slashed{D} = \gamma^\mu D_\mu = \gamma^\mu (\partial_\mu - i e A_\mu)$, and $F_{\mu\nu} = \partial_\mu A_\nu - \partial_\nu A_\mu$ is the field strength tensor. The fermionic functional integral can be performed exactly due to the simple structure of the Dirac operator in lower dimensions.

It is straightforward to see that, in our generalised framework with the regularised measure $\Dcal (\Rcal_{\eta} \psi(x)) \Dcal (\Rcal_\eta^{\dagger} \bar{\psi}(x))$, we obtain the regularised Jacobian under a $2D$ chiral transformation \citep{Fujikawa:2004cx}

\be\label{2DRegJacobian}
\ln \left(\Rcal_\eta J_5(\alpha) \Rcal_\eta^{\dagger} \right) = -2i \sum \limits_{n=1} \int d^2 x \, \phi_n^{\dagger}(x)\alpha(x) \gamma_5 \eta(\Ocal)\phi_n(x).
\ee

For the choice of Fujikawa's cut-off $\eta(\Ocal) = \exp \left(-\slashed{D}^2/\Lambda^2\right)$ the chiral anomaly is given as

\be\label{2Dchiralanomaly}
\Tr(\Rcal_\eta \gamma_5 \Rcal_\eta^{\dagger}) = \frac{i}{2\pi} \int d^2x \, \epsilon^{\mu \nu}F_{\mu \nu},
\ee
agreeing with established results \citep{Fujikawa:2004cx}. Here, $\eta(x)$ is once again a smooth regulator satisfying the properties $\eta(0)=1$ with rapid decay as $x \to \infty$. 

The key point, returning to the regularised Jacobian \eqref{2DRegJacobian}, is that we can rewrite this expression for the regularised sum over modes directly in terms of the heat kernel. From the spectral decomposition, one may instead choose the heat kernel cut-off $\eta(\Ocal) = \exp(-s \lambda^2_n)$ so that

\be\label{HeatKernelTr}
\text{Tr}(\gamma_5 \eta(\Ocal)) = \lim_{s \to 0} \sum_n \phi_n^\dagger \gamma_5 e^{-s D^2} \phi_n,
\ee
in which the parameter $s$ is the proper time. In the proper time representation, Fujikawa's regularisation just amounts to setting $s \sim 1/\Lambda^2$ \citep{Umezawa:1989ns}. One can also see this directly from the regularised functional trace of the form $\Tr(\gamma_5 \exp(-\slashed{D}^2/\Lambda^2))$, where in the naive transformation to Schwinger proper time representation the regulator takes the form of a step function $\theta(s - 1/\Lambda^2)$ that picks out exactly one value of $s = 1/\Lambda^2$ in the proper time integration. The reason for this will become clear later on.

From \eqref{HeatKernelTr}, one can transform the expression for the regularised trace into the heat kernel representation

\be\label{HeatKernelTr2}
\text{Tr}(\gamma_5 \eta(\Ocal)) = \lim_{s \to 0} \int d^2x \, \text{Tr} \left[\gamma_5 K(x,x,s) \right],
\ee
where $K(x, x'; s)$ is the heat kernel associated with $\slashed{D}$ satisfying

\be
\left( \frac{\partial}{\partial s} + \slashed{D}^2 \right) K(x, x^{\prime}; s) = 0, \quad K(x, x^{\prime}, 0) = \delta^2(x - x^{\prime}).
\ee

In flat two dimensional Euclidean space, the heat kernel has a well-known asymptotic expansion for small $s$ in terms of Seeley-DeWitt coefficients \citep{Ball:1988xg}. This is given by

\be
K(x,x;s) = \frac{1}{(4\pi s)} \sum \limits_{n=0}^{\infty} s^n a_n(x,x),
\ee
where $a_n(x,x^{\prime})$ are the Seeley-DeWitt coefficients that encode the local geometric properties the operator $\slashed{D}^2$. The leading term $a_0(x)$ is proportional to the identity, while the next term is $a_1(x, x)$ is proportional to $ F_{\mu \nu}$ containing the first non-trivial contributions of the gauge field. Since $\tr(\gamma_5 \sigma^{\mu\nu}) = \epsilon^{\mu\nu}$, we recover the chiral anomaly \eqref{2Dchiralanomaly}. In fact, we may define the anomaly function in two dimensions explicitly as the limiting value of the heat kernel  

\be
\mathcal{A}(x) = \lim_{s \to 0} \text{Tr} \left[ \gamma_5 K(x, x; s) \right].
\ee

To make better sense of these observations and to draw deeper connections, let us now compare with the generalised Schwinger proper time approach developed in \citep{Ball:1988xg}, which is  inspired by the heat kernel formalism. From the theory of massless Dirac fermions \eqref{SchwingerLagragian}, one can instead consider the effective action written as a determinant of the Dirac operator

\be
e^{-\Gamma_{\text{eff}}[A]} = \int [\mathcal{D}\bar{\psi} \mathcal{D}\psi] e^{-\int d^2x \bar{\psi} (\slashed{D}) \psi} = \det(\slashed{D}),
\ee
which, upon taking the logarithm, gives the standard fermionic determinant in two dimensions

\be
\Gamma_{\text{eff}}[A] = - \ln \det (\slashed{D}).
\ee

Using the identity $\ln \det (\slashed{D}) = \frac{1}{2} \text{Tr} \ln (\slashed{D}^2)$, one can transform this expression for the effective action directly to the generalised Schwinger proper time representation such that

\be\label{EffactionFunctdet}
-\ln \det (\slashed{D}) = -\frac{1}{2} \int_0^\infty \frac{ds}{s} \int d^2x \operatorname{Tr} K(x, x; s),
\ee

where we note $(\slashed{D})^2 = \bigl(D^2 + \frac{1}{2} \,\sigma_{\mu\nu}F^{\mu\nu}\bigr)$ with $\sigma_{\mu\nu} = \frac{1}{2}[\gamma_\mu,\gamma_\nu]$. In this setting, the chiral anomaly is seen to be associated with a variation of the effective action. In the heat kernel representation, one in fact writes
\be\label{EffactionFunctdet2}
\delta \Gamma_{\text{eff}}[A] = -\frac{1}{2} \int_0^\infty \frac{ds}{s} \int d^2x \, \alpha(x) \operatorname{Tr}(\gamma^5 K(x, x; s)),
\ee
where the trace is defined over Dirac and internal indices. Compared with \eqref{HeatKernelTr2}, the main difference is that the UV divergence has been shifted completely to the proper time variable $s$. This means the $s$ integration requires regularisation. Moreover, the general integral structure in \eqref{EffactionFunctdet2} for anomaly calculations in proper time is based on the definition of a general family of Schwinger proper time integrals. This family of integrals, one for each $s \in [0, \infty)$, is given by

\be\label{FunctdetGenSchwing}
= -\frac{1}{2} \int_0^\infty \frac{ds}{s} \rho(\Lambda,s) \, \int d^2x \, \alpha(x) \operatorname{Tr}(\gamma^5 K(x, x; s)),
\ee
with $\rho(\Lambda,s)$ a class of smooth regulator functions that regularises the proper time integration, ensuring convergence while preserving gauge invariance \citep{Ball:1988xg}. This regulator $\rho(\Lambda, s)$ is not as arbitrary as it may first appear. As we'll see, it is closely related to the class of $\eta$ regulators \citep{PadillaSmith1} derived from irreducible loop integrals, and, consequentially, with the regularised measure in the path integral. Just the same, $\rho(\Lambda, s)$ must satisfy $\rho(s) \to 1$ for large $s$, and $\rho(s) \to 0$ as $s \to 0$. Common choices of $\rho$ including proper time cut-off, Pauli-Villars, dimensional regularisation, and zeta function regularisation. 

Given we are interested in a conceptual comparison, it is notable that equation \eqref{HeatKernelTr2} gives a regulated trace arising directly from the path integral measure (Fujikawa’s approach), while the expression for $ \delta \Gamma_{\text{eff}}$ in \eqref{FunctdetGenSchwing} represents the anomaly as a functional variation of the effective action, requiring integration over the proper time parameter $s$ and insertion of the local chiral parameter $ \alpha(x)$. Despite the structural differences, both formulations involve the same spectral content via $\Tr(\gamma^5 K(x,x;s))$, and the regulator $\rho(\Lambda,s)$ provides a unified smoothing between these approaches.

The deeper point is that, as observed from \eqref{HeatKernelTr}, Fujikawa's exponential regularisation, which suppresses high eigenmodes in the spectral representation, can be approximated for the choice $\rho(\Lambda, s) = \theta(s - 1/\Lambda^2)$ in the Schwinger proper time representation \citep{Ball:1988xg}. In Schwinger proper time, one defines the Green's function or effective action so that short time contributions $s \leq 1/\Lambda^2$ are not integrated over, or are effectively damped by $\rho(\Lambda, s)$. The difference in the two representations of the Fujikawa cut-off function is just the difference in the spectral and proper time descriptions of the same regularised theory.

Tying all of these approaches together, we note that from a momentum space perspective, the Euclidean heat kernel $\exp(-sD^2)$ can be expanded directly in terms of momentum eigenstates. For simplicity, from the fermionic determinant \eqref{EffactionFunctdet}, using the plane wave basis one finds $\langle x| \exp(-sD^2) |x \rangle = \exp(-s(k^2 + M^2))$. Hence, one can transform the position space integral to momentum space

\be
\Gamma_{\text{eff}}^{Reg}[A]= -\frac{1}{2} \int d^2x \, \int_0^\infty \frac{ds}{s}  \rho(\Lambda, s) \, \int \frac{d^2k}{(2\pi)^2} e^{-s(k^2 + M^2)}.
\ee

The result is precisely the representation of $\eta$ regularisation in generalised Schwinger proper time \citep{PadillaSmith1}. While it is known that a regulator $\eta(k^2/\Lambda^2)$ can be represented in Schwinger proper time form under certain limits---particularly $M^2/\Lambda^2 \to 0$ as emphasized in \citep{PadillaSmith1}---our aim here is to highlight the deeper structural unification of these schemes. Namely, that the same divergent trace can be regularised consistently via $\rho(s)$ in proper time, $\eta(\lambda^2)$ in spectral space, or $\eta(k^2)$ in momentum space, all preserving gauge invariance and yielding the same chiral anomaly in the Schwinger model.

Indeed, from the spectral analysis the generalised Schwinger proper time regulator $\rho(\Lambda, s)$ can be related directly to the $\eta$ regulators in the momentum space representation \cite{PadillaSmith1} via a modified or weighted Laplace transform. (If $M^2 \ll \Lambda^2$, then the Laplace equivalence is only approximate). This draws further connection between Fujikawa's spectral cut-off, generalised Schwinger proper time, and $\eta$ regularisation of loop integrals. One furthermore sees that the eigenvalue sum discussed in Section \ref{EigensumSeries} can be written in the form $\sum \limits_{n}^{\infty} \langle n| \gamma_5 |n \rangle \eta(\Lambda, \lambda_n)$. For the choice of $\eta(\Lambda, \lambda_n)$ equivalent to Fujikawa's regulator, we obtain the same asymptotic value of the regularised sum in \eqref{FujikawaDivergentSumReg2}. And since $\eta$ is equivalent to $\rho$ in their respective representations, we once again observe a unifying picture of how the same divergent series is regularised with the requirements of maintaining gauge invariance, with the resulting spectral asymmetry in the regularised summation ultimately leading to the chiral anomaly.

Of course it is trivial to observe that one can regularise the same quantity in three different ways; but it would be interesting to study more deeply the possible equivalence of these different albeit related regularisation prescriptions. Furthermore, it is intriguing to ask whether there is a mapping between corresponding regulators, and whether any such mapping might be insightful in terms of a general theory of regularisation.

\section{Conclusions}
In this paper, we have explored the relationship between Fujikawa’s path integral formulation of the chiral anomaly and the more general idea of the regularised functional measure. Beginning with a review of Fujikawa’s formalism, we emphasised the central role played by the divergent sum over eigenmodes of the Dirac operator in the computation of the anomaly. This sum, appearing in the functional Jacobian transformation under chiral symmetry, is inherently ill-defined, requiring an appropriate regularisation scheme to extract meaningful physical results. 

To understand the mathematical structure of this divergence, we analysed the spectral summation in detail, demonstrating that the sum over eigenmodes takes the form of a generalised alternating series. Using smoothed asymptotics, we rigorously computed the weighted sum over eigenmodes, making use of the smoothed summation method associated with $\eta$ regularisation to ensure well-defined convergence. This analysis not only reinforced the argument of regularising the functional trace in the sense of how it is defined, but also highlighted the interplay between spectral asymmetry and anomaly generation.

Building upon this foundation, we then formally defined a generalised framework for defining the regularised measure, incorporating an operator-valued regulating function that naturally captures Fujikawa’s original expression for the chiral anomaly from the outset. This framework provided a rigorous justification for the regularisation of the path integral measure and allowed us to recover the standard anomaly result through a properly regulated spectral trace. We further derived the $\iota_{E}(\Lambda)$ function, encoding the response of the regularised measure to changes in the cutoff parameter.

Our findings demonstrate that the regularisation of the path integral measure is not an ad hoc procedure, but rather an intrinsic feature in an approach to QFT that ensures well-defined anomaly computations. The identification of the Mellin transforms as a key feature of anomaly regularisation further suggests deeper connections to renormalisation theory and asymptotic analysis. 
 
In the final part of this work, we commented on conceptual connections to the Schwinger proper-time formalism, reviewing the direct correspondence between the regularised measure, proper-time cutoff methods, and $\eta$ regularisation for loop integrals. We discussed in what way Fujikawa’s exponential suppression of high-energy eigenmodes corresponds, in the proper-time framework, to an integration cutoff in the short-time region. By introducing a smooth proper-time regulator function $\rho(\Lambda,s)$, we surveyed how the regularisation of the path integral measure translates into a well-defined regularisation prescription within the heat kernel expansion. Finally, by transforming the regulated proper-time integral into momentum space, we made explicit the equivalence between $\eta$ regularisation and proper-time regularisation, establishing that both methods belong to a broader class of gauge-invariant regulators for anomaly calculations that includes Fujikawa's regularisation scheme. This reveals the edges of a unified picture between different representations including Fujikawa's choice of cutoff, spectral regularisation, and the more general framework of $\eta$ regularisation.

Future directions include extending these techniques to higher-dimensional anomalies, investigating their implications in topological field theories, and applying the regularised functional measure to gauge-invariant computations in curved spacetime. Of particular interest is the prospect of exploring the non-perturbative renormalisation of QED within this framework. Additionally, our framework may offer a new viewpoint on the multiplicative anomaly, a subtle feature arising in the regularisation of operator products, typically computed using standard zeta or heat kernel techniques. Further investigation could clarify whether $\eta$ regularisation, in conjunction with the functional measure, provides a consistent scheme for regularising products of elliptic operators in curved backgrounds.

More broadly, a full generalisation of our framework to curved spacetime naturally raises several questions related to UV completion. As discussed in Section \ref{SecIota}, if the theory is interpreted as an effective field theory, or if background fields such as the metric are eventually promoted to dynamical variables, then previous simplifications no longer apply. In such cases, divergences that appear to be removable in fixed-background settings---such as constant contributions to the regularised measure---can become field-dependent, thereby affecting the relative weighting of configurations in the full path integral. This suggests that a fully regulator-independent measure, as encoded in the vanishing of all relevant Mellin transforms, may be essential for a consistent UV completion---provided that a suitable definition of super-enhanced regulators can be formulated to ensure such vanishing even after differentiation.

An open question remains as to whether residual divergences, even when reduced to constant factors for fixed backgrounds, can still signal a breakdown of convergence when background fields are allowed to vary. If effective descriptions yield partition functions that differ only by constant prefactors for each fixed background configuration, does this guarantee a convergent path integral when all configurations are summed over? Or might such prefactors encode non-trivial global obstructions to UV completion?

Finally, we note recent developments by Branchina et al. \citep{Branchina:2024xzh,Branchina:2024lai}, which investigate a proper treatment of the functional measure in the context of the cosmological constant problem. Exploring potential connections between their approach and the regularised measure formalism proposed here may yield new insights. Similarly, Bilal and Ferrari \citep{Bilal:2013iva} applies smooth spectral cutoff methods to link zeta-function regularisation with spectral flow techniques in curved spacetime, offering another possible point of contact. We also note that, from the perspective of our generalised framework, it would be valuable to understand whether these connections can be understood in a scheme-independent way similar to \citep{PadillaSmith2}.

\section*{Acknowledgements}
We are grateful to Tony Padilla and Peter Millington for useful comments and discussions. RGCS was supported by a Bell Burnell Studentship. For the purpose of open access, the authors have applied a CC BY public copyright licence to any Author Accepted Manuscript version arising. No new data were created during this study.

\bibliography{Nonpertrenormrefs}
\end{document}